# Ultra-high endurance silicon photonic memory using vanadium dioxide

**Juan José Seoane[1], Jorge Parra[1], Juan Navarro-Arenas[1,2], María Recaman[3], Koen Schouteden[3], Jean Pierre Locquet[3] and Pablo Sanchis[1*]**

[1]*Nanophotonics Technology Center, Universitat Politècnica de València, Camino de Vera s/n, 46022, Valencia, Spain Department*

[2]*Institute of Materials Science (ICMUV), Universitat de València, Carrer del Catedràtic José Beltrán Martinez 2, 46980, Valencia, Spain*

[3]*Department of Physics and Astronomy, KU Leuven, Celestijnenlaan 200D, 3001 Leuven, Belgium*

*\*Corresponding author:* pabsanki@ntc.upv.es

## Abstract

Silicon photonics arises as a viable solution to address the stringent resource demands of emergent technologies, such as neural networks. Within this framework, photonic memories are fundamental building blocks of photonic integrated circuits that have not yet found a standardized solution due to several trade-offs among different metrics such as energy consumption, speed, footprint, or fabrication complexity, to name a few. In particular, a photonic memory exhibiting ultra-high endurance performance (> $10^6$ cycles) has been elusive to date. Here, we report an ultra-high endurance silicon photonic volatile memory using vanadium dioxide ($VO_2$) exhibiting a record cyclability of up to $10^7$ cycles without degradation. Moreover, our memory features an ultra-compact footprint below 5 μm with the potential for nanosecond and picojoule programming performance. Our silicon photonic memory could find application in emerging photonic applications demanding a high number of memory updates, such as photonic neural networks with in-situ training.

## Introduction

Silicon photonics has open new opportunities to meet the growing computational and data communications demands in today's world by offering a cost-effective and scalable platform[1]. However, the photonic memory is still a missing building block in current silicon platforms. While there are several technologies that may be used for enabling photonic memories, to the best of our knowledge, this component has not yet been included in any Process Design Kit (PDK) of CMOS foundries offering either Multi-Wafer Project (MWP) runs or custom fabrication. Such a device may be essential for numerous applications but particularly in the emerging fields of neuromorphic computing and programmable photonics [2]. In these scenarios, the number of cycles an integrated photonic memory can undergo could be a critical metric for applications requiring high-cyclability, such as in-situ neural network training [3,4].

Over the past few years, various technologies, mainly phase change materials (PCMs), ferroelectrics and charge-trapping memories, have been explored for implementing photonic memories[5,6]. PCMs stand out for a dramatic change in their refractive index between material phase states, enabling devices with ultra-compact footprint and thereby showing a high potential for large-scale integration. Furthermore, most of the PCMs considered for photonics could be monolithically integrated in silicon platforms.

Chalcogenide-based photonic memories have arisen as a promising approach, primarily due to their non-volatile switching capability, which has been used for developing different photonic in-memory computing architectures[7–12]. However, one of the limitations of chalcogenides is the relatively poor endurance performance that has been demonstrated in hybrid photonic integrated devices, from typically $\sim 10^3$ switching cycles[6] up to 20,000 cycles[13].

As an alternative to chalcogenides, vanadium dioxide ($VO_2$) has the potential to convert into an appealing candidate to build photonic memories with an ultra-compact footprint by exploiting the hysteretic response of its refractive index variation with temperature[14]. $VO_2$ could pave the way for faster and more energy-efficient devices due to its much lower switching temperature compared to chalcogenides, although at the expense of a volatile nature[15]. Nevertheless, the possibility for a non-volatile switching behavior at room temperature has also been recently reported [16]. In this work, we report a high-endurance ultra-compact $VO_2$/Si photonic volatile memory showing a record cyclability of up to $10^7$ write/erase cycles with speed and energy consumption outperforming chalcogenide-based non-volatile photonic memories [5,6]. The volatile nature of our proposal makes it suitable for applications where frequent switching is required, rather than long-term storage applications.

The proposed memory is shown in Fig. 1(a), and it comprises a standard 220x500 nm silicon waveguide loaded with a 3 µm-long and 40-nm-thick $VO_2$ patch deposited by molecular-beam epitaxy (MBE) [15]. The device works at 1550 nm and for the transverse magnetic (TM) polarization. The memory is programmed by using optical pulses to photothermally drive the $VO_2$ between a low-loss insulating phase (erased memory state) and a high-loss metallic phase (written memory state). The simulated performance of the device is shown in the supplementary note 1. Figure 1(b) shows an optical image of the fabricated device. To hold the memory state, the optical power of the programming pulses is adjusted to fall within the hysteresis loop of the $VO_2$ metal-to-insulator transition, as depicted in Fig. 1(c). Nevertheless, microheaters could also be used to provide the holding temperature.

## Results

### Endurance measurements

To demonstrate the maximum write/erase cycles our memory can undergo before it shows degradation signals, we repeated the write/erase cycle in the programming signal, illustrated in Fig. 1, with a period of 100 µs. The period was chosen so it was the lowest possible using our set-up to achieve the required optical power to operate the memory. We took a 5-cycle trace every second, which corresponds to a difference among traces of $10^4$ cycles. Both write and erase optical pulses had a duration of 1 µs, while the state of the memory was held by using just 180 µW.

Figure 2 shows the cyclability tests for evaluating the endurance performance with an optical contrast of ~2.6 dB between the written/erased states. The memory operation was successfully proven over $10^6$ cycles [Fig. 2(a)-(c)]. Above this value, the optical contrast was gradually decreased due to optical misalignment and drifts in the experimental set-up. However, after realignment, the optical contrast between states was recovered, as shown in Fig. 2(a), thus demonstrating a record value exceeding $10^7$ cycles. It is important to notice that even after this huge number of cycles, the device continued working correctly. Moreover, the distribution of the optical transmission readout, shown in Fig. 2(d), revealed a remarkable accuracy of ±0.11 dB in the erased state and ±0.085 dB in the written state for the standard deviation.

**Speed and energy consumption performance**

In order to evaluate the potential programming speed and energy consumption, we applied a single optical pulse with the shortest duration (~100 ns) available by our setup into a smaller photonic memory of just 1 μm length. Figure 3 shows the transmission performance when injecting optical pulses with increasing energy. The switching time is reduced from 28 to 12 ns for the $VO_2$ insulator-metal phase transition (writing process) when the energy pulse increases but at the expense of a longer switching time for the $VO_2$ metal-insulator phase transition (erasing process) that increases from 36 to 208 ns. Hence, an energy pulse of 24 pJ was sufficient to switch back and forth the $VO_2$ patch with switching times as fast as 24 and 36 ns, respectively.

Figure 4 depicts a comparison between different technologies employed for photonic memories such as chalcogenides, memristors, and charge-trapping, in terms of programming energy consumption and endurance. More details of the compared works are detailed in the supplementary note 2. Our $VO_2$/Si memory approaches the ultra-low energy consumption of memristor technology but showcases four orders of magnitude increase in endurance. On the other hand, compared to the most used chalcogenide employed in photonic memories, $Ge_2Sb_2Te_5$ (GST), our experimental results lead to a switching energy density of 0.4 aJ $nm^{-3}$ to write the state, which is one order of magnitude lower than experimental state-of-the-art GST photonic memories (8 aJ $nm^{-3}$) and even lower than the theoretical limit of GST (1.2 aJ $nm^{-3}$)[17].

Following the same procedure as in reference[17], we can estimate the theoretical energy density limit to write the state in VO$_2$. Considering the latent heat,[18,19] $H = 235 \, J \, cm^{-3}$, the specific heat capacity[20], $C = 3 \, J \, cm^{-3} K^{-1}$ and the temperature where the complete transition to metal occurs T = 65 ºC, we obtain a programing energy density of $H + C \, \Delta T = 0.31 \, aJ \, nm^{-3}$, where $\Delta T = 65 - 25 \, °C$. However, in our memory device, such a temperature increase is lower since the programming process occurs from the holding temperature instead of the room temperature. Therefore, latent heat is only required to finish the transition and the theoretical energy density for programming a VO$_2$ memory would be $H = 0.23$ aJ nm$^{-3}$, which is near to our experimental value.

## Discussion

In this work we have reported an ultra-compact VO$_2$/Si photonic memory with a record endurance of up to 10$^7$ cycles. Furthermore, we have also demonstrated the potential for operating the memory with speeds of a few nanoseconds and energy consumption of a few picojoules. This result implies a huge increase in endurance from previous works and a significant reduction in programming energy consumption. To contextualize these values, in an application where the memory is erased and written with a frequency in the range of kHz, current photonic memories based on chalcogenides would only endure a few seconds before they experience a functionality degradation[8,29]. Moreover, most chalcogenide-based photonic hardware employed for neural network applications uses the offline learning approach[4]. In this procedure, the training of the neural network is done in the usual manner, computing the weights and biases via a backpropagation algorithm in a computer[30]. After that, the weights and biases are written onto the chalcogenide patches. Although this method is very useful in several cases, it does not offer any energy consumption or speed enhancement throughout the entire training process. Moreover, fabrication imperfections may imply deviations from the original (digital) parameters, resulting in a reduction of accuracy from the simulated model, the so-called 'reality gap'[31]. In contraposition to this approach, in the *in situ training*[32] or online training[4], the photonic hardware is used to carry out the training. This approach would imply a substantial energy consumption reduction of neural network training, external computing tasks would be minimized, and fabrication imperfections would not affect the transfer of the digital weights to the analogic weights in the hardware (the 'reality gap' would not be present). In this regard, $VO_2$ is positioned as a potential candidate for in situ training thanks to its high endurance and low programmable energy requirements. To summarize, our device could offer a promising solution for applications requiring memory functionality with high-cyclability together with low-power and fast-speed operation.

## Methods

### Experimental set-up

A contra-directional pump and probe technique was used to carry out both the endurance measurements and the energy and speed performance. A low-power continuous wave signal at 1565 nm was used to readout the change of the memory, whereas an externally modulated signal at 1550 nm was employed for programming and holding the state of the memory. The programming signal was modulated via an electro-optical modulator and amplified using an erbium-doped fiber amplifier (EDFA). After pump (programming signal) and probe (readout) signals propagated through the sample, high-speed photodetectors were used to measure them with an oscilloscope. Moreover, one additional photodetector was used to obtain the chip transmission for real-time monitoring. Both input and output fibers were manually aligned, and the light was coupled onto the chip through grating couplers.

**Fabrication process**

The silicon photonic structures were fabricated on a standard silicon-on-insulator (SOI) sample with a top layer of 220-nm-thick silicon with a 3-µm-thick buried oxide layer. E-beam lithography was employed to pattern the silicon structures onto a negative tone resist. The patterning was transferred into the SOI sample by employing plasma-reactive ion etching (ICP-RIE). The $VO_2$ structures were defined using the same e-beam lithography process onto a positive tone resist, with a subsequent development of the exposed areas using an MBIK:IPA bath. Then, a 40-nm-thick $VO_x$ layer was grown using molecular beam epitaxy (MBE), followed by a lift-off process using MBIK:IPA in an ultrasonic bath. Then, polycrystalline $VO_2$ was formed by carrying out an annealing in forming gas at 450 ºC for 30 min. Finally, the $VO_2$/Si structures were covered with a 700-nm-thick $SiO_2$ cladding deposited using plasma-enhanced chemical vapor deposition (PECVD) at 200 ºC.

**Data availability.** Data are available upon reasonable request.

**Acknowledgments.** The authors would like to acknowledge Dr. Amadeu Griol and Dr. Dora Ivanova for their work in the fabrication of the devices at the Nanophotonics Technology Center (Universitat Politecnica de Valencia, Spain). This work was supported by funding from the European Union's Horizon Europe by the granting Authority "HADEA (European Health and Digital Executive Agency) under Grant Agreement No 101070690 (PHOENIX). Grant PID2022-137787OB-I00 funded by MCIN/AEI/10.13039/501100011033 and by "ERDF A way of making Europe", and PROMETEO Program (CIPROM/2022/14) funded by Generalitat Valenciana are also acknowledged. J. Navarro acknowledges grant from University of Valencia/Ministry of Universities (Government of Spain), modality "Margarita Salas" (MS21-037), funded by the European Union, Next-Generation EU. J. Parra would like to acknowledge Universitat Politècnica de València for his grant 2-PAID-10-22.

**Author Contributions.** J.J.S., J.P. and P.S wrote the main manuscript. M.R., K.S. and J.P.L. fabricated the $VO_2$ patches. J.J.S. carried out the measurements with the assistance of J.P. Results were analyzed by J.J.S., J.P., J.N-A. and P.S. All authors reviewed and approved the manuscript.

**Competing Interests.** The authors declare no conflicts of interest.

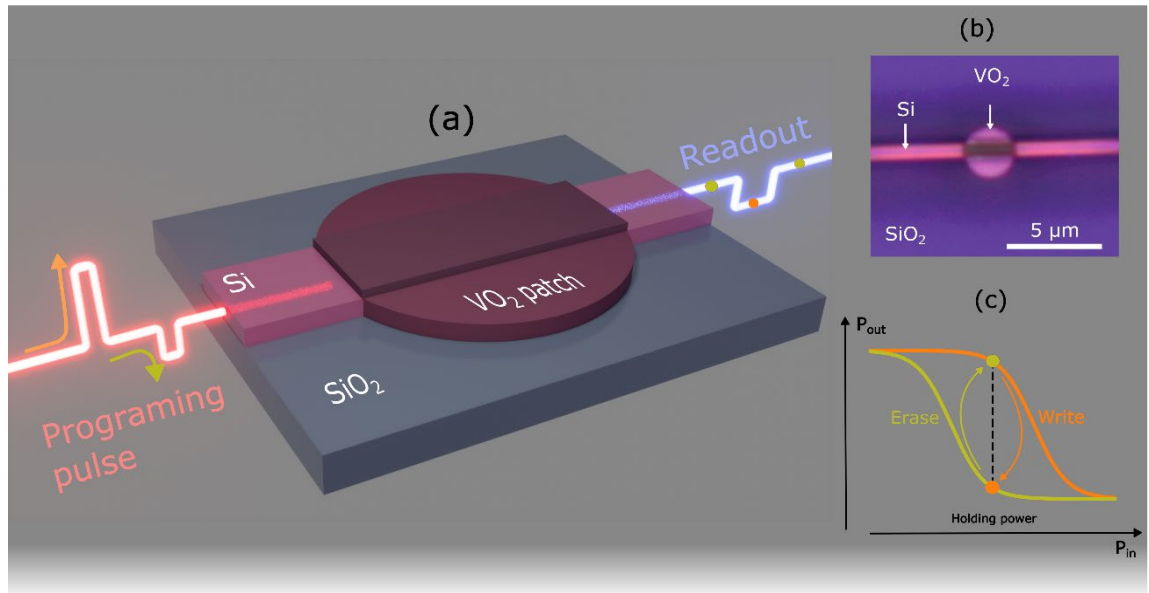

**Figure 1**. **Device scheme and working principle**. (a) Working principle of our VO$_2$/Si photonic memory operated by programming (write/erase) optical pulses. (b) Optical image of the fabricated device. (c) Sketch of the memory operation between written/erased states by exploiting the hysteretic response of the VO$_2$ insulating-to-metal phase transition.

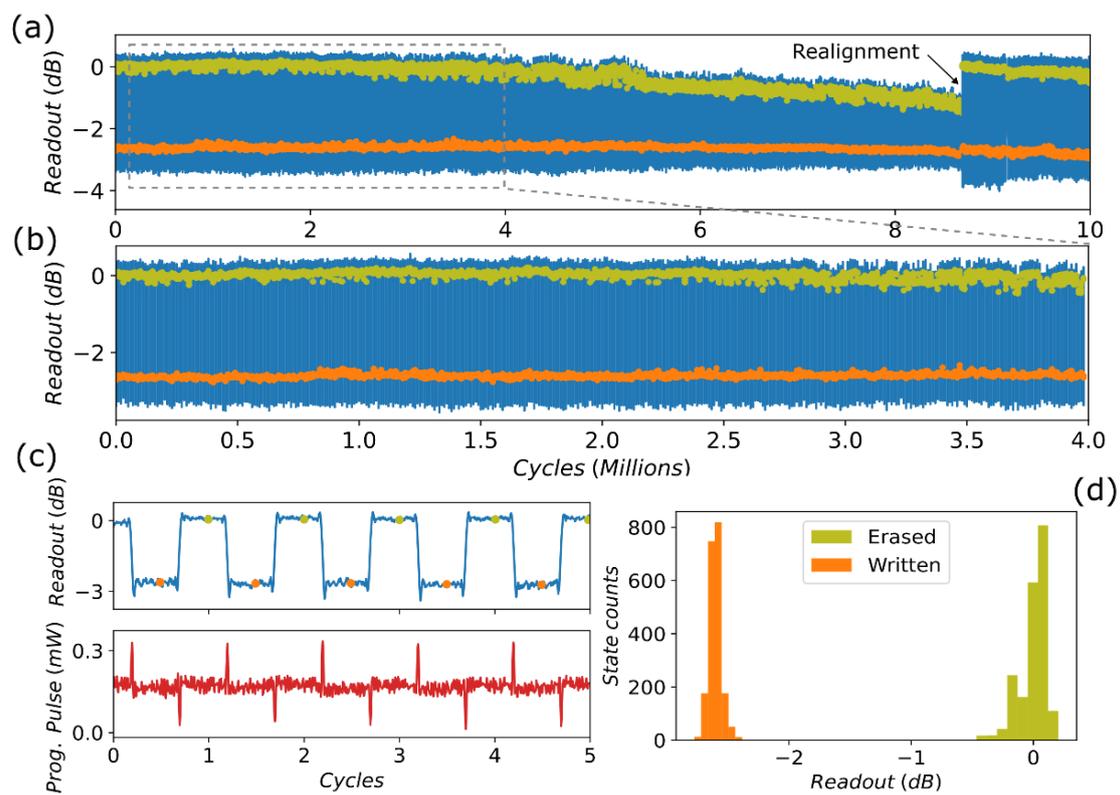

**Figure 2 Endurance measurements record**. (a) Memory operation demonstrating more than a million cycles and just limited by misalignment. (b) Stable operation over $10^6$ cycles before misalignment. (c) Detailed view of one of the traces showing 5 write/erase cycles with its corresponding programming pulse. (d) Distribution of the optical transmission readout after $4\times10^6$ cycles.

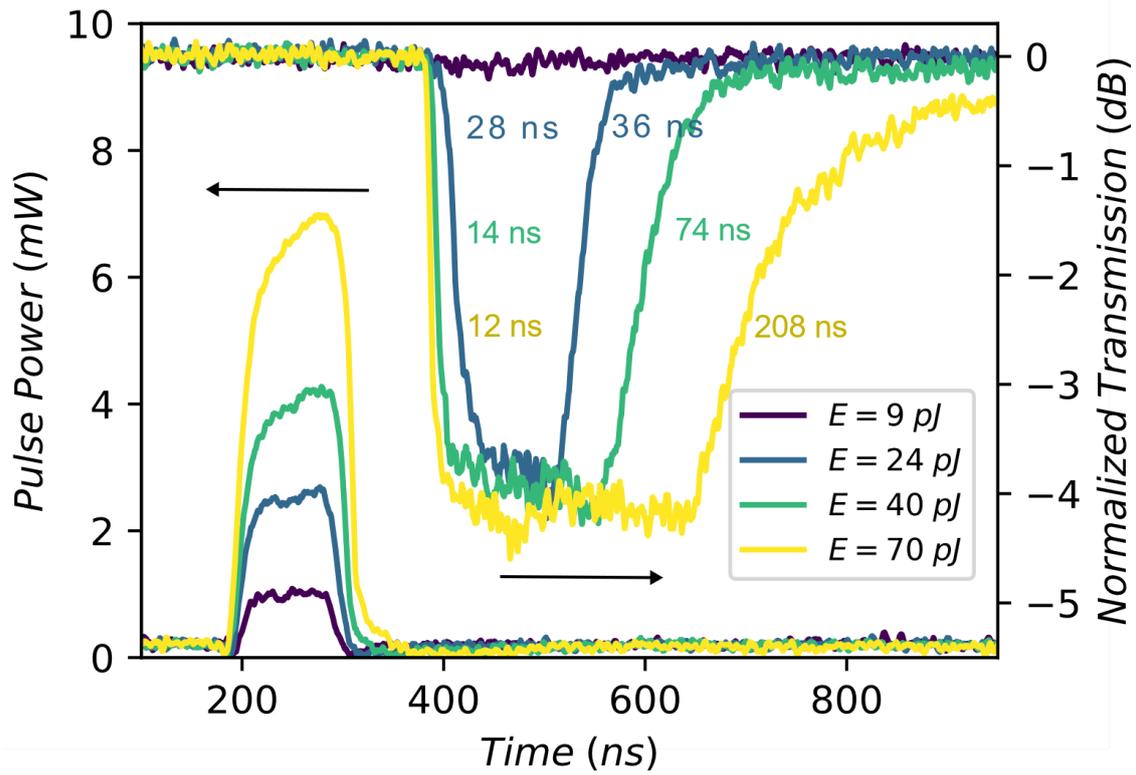

**Figure 3. Switching speed and energy consumption performance.** Switching operation by applying 100 ns-long optical pulses with increasing energies into smaller photonic memory of just 1 µm length. The ~200 ns delay between the injecting pulse and output response is mainly caused by the external components and fiber used in the setup.

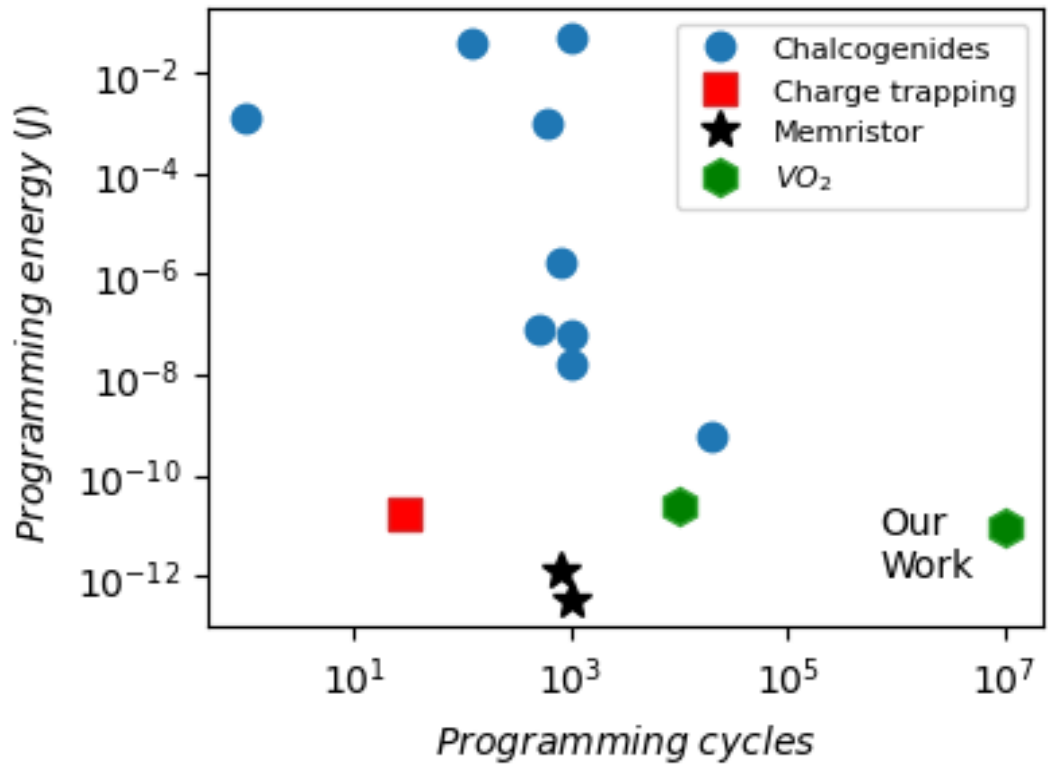

**Figure 4. Programming energy vs endurance comparison.** Programming energy consumption vs endurance of our work compared with other technologies such as chalcogenides[6,13,21,21–26,30] (blue circle), charge-trapping memories[27] (red square) memristors[6,28] (black star) and other $VO_2$ memories[14] (green hexagon).